\documentclass[11pt]{article}

\usepackage{scicite}
\usepackage[T1]{fontenc}
\usepackage{ragged2e}

\usepackage{times}
\usepackage{soul}

\newenvironment{sciabstract}{%
\begin{quote} \bf}
{\end{quote}}

\title{Noise Resistant Phase Imaging with Intensity Correlation}

\author
{Jerzy Szuniewicz$^{1}$, Stanis\l{}aw Kurdzia\l{}ek$^{1}$, Sanjukta Kundu$^{1}$, Wojciech Zwolinski$^{1}$,\\ Rados\l{}aw
Chrapkiewicz$^{2}$, Mayukh Lahiri$^{3}$, Radek Lapkiewicz$^{1\ast}$ \\
\\
\normalsize{$^{1}$Institute of Experimental Physics, Faculty of Physics, University of Warsaw,}\\
\normalsize {ul. Pasteura 5, 02-093 Warszawa, Poland,}\\
\normalsize{$^{2}$CNC Program, Stanford University, Palo Alto, CA 94304, United States}\\
\normalsize{$^{3}$Department of Physics, Oklahoma State University, Stillwater, OK 74078-3072, United States}\\
\normalsize{$^\ast$radek.lapkiewicz@fuw.edu.pl}
}

\date{}

\usepackage{setspace}

\usepackage{lipsum}

\usepackage{xcolor}
\usepackage{graphicx}

\setstcolor{teal}

\begin{document} 

\singlespacing



\maketitle


\begin{sciabstract}
Interferometric methods, renowned for their reliability and precision, play a key role in phase imaging. Interferometry typically requires high coherence and stability between the measured and reference beams. The presence of rapid phase fluctuations washes out interference fringes, making phase profile recovery impossible. This difficulty can be circumvented by significantly shortening the measurement time. However, such an approach results in reduced photon-counting rates, precluding applications in low-intensity imaging. We introduce and experimentally demonstrate a phase imaging technique that is immune to time-dependent phase fluctuations. Our technique differs fundamentally from conventional phase imaging techniques since it relies on intensity correlation instead of direct intensity measurements. This allows one to obtain high interference visibility for arbitrarily long acquisition times and is therefore particularly advantageous when the photon flux is low. We prove the optimality of our method using the Cram\'er-Rao bound in the extreme case when no more than two photons are detected within the time window of phase stability. Our technique will open new prospects in phase imaging, including emerging applications such as in infrared and X-ray imaging, and quantum  and matter-wave interferometry.
 
\end{sciabstract}

\section*{Introduction}
In the field of quantitative phase imaging\cite{Cuche1999,Ikeda2005,PopescuBOOK}, interferometric methods play a crucial role in capturing phase information of a sample through the measurement of phase shifts\cite{mertz_2019}. These phase shifts can provide insights into the refractive index, thickness, and structure of the sample. Consequently, phase imaging has made a significant impact on various fields such as bio-medical imaging\cite{Park2018,Popescu2006,Wang2011}, sample-safe defectoscopy\cite{Makowski2022}, wavefront sensing\cite{Schnars2015},  and 3D holography\cite{Choi2007}. Interferometry\cite{interference_book} is central to phase imaging due to its ability to detect minute variations in optical paths. Moreover interferometry is the basis of optical coherence tomography\cite{Huang1991,Sticker2001}. Some of the interferometric techniques, especially those related to biology, require very low photon fluxes.
For an interferometric measurement a wave field that has interacted with an object is superposed with a reference field and the resulting interference pattern is detected by a camera. If the object and the reference fields are mutually coherent, the time-averaged intensity on the camera is given by\cite{book_Hecht, PopescuBOOK}:
\begin{equation}
I(x,y)=I_{\mathrm{r}}+I_{\mathrm{o}}+2\sqrt{I_{\mathrm{r}} I_{\mathrm{o}}}\cos[\Theta + \phi(x,y)],
\label{eq1}
\end{equation}
where $I_{\mathrm{r}}$ and $I_{\mathrm{o}}$ are intensities of the reference and the object fields respectively, $\Theta$ is the spatially uniform phase that can be changed by introducing delay between the two fields, and $\phi(x,y)$ is the phase profile of the object. Standard interferometric phase imaging techniques are based on the signature of $\phi(x,y)$ left in the detected intensity pattern. However, for any such method to be applicable, the object and the reference fields need to be mutually coherent and phase $\Theta$ must be stable in time. The method is therefore vulnerable to time-dependent, uncontrollable phase fluctuations. 
When the phase fluctuates much faster compared to the detection time, the coherence between the object and image fields is effectively lost, and no interference is observed, i.e.,
\begin{equation}
I(x,y)=I_{\mathrm{r}}+I_{\mathrm{o}}.  
\label{inten-incoh}
\end{equation}
\begin{figure}[h!]
   \centering
 \includegraphics[width=0.9\linewidth]{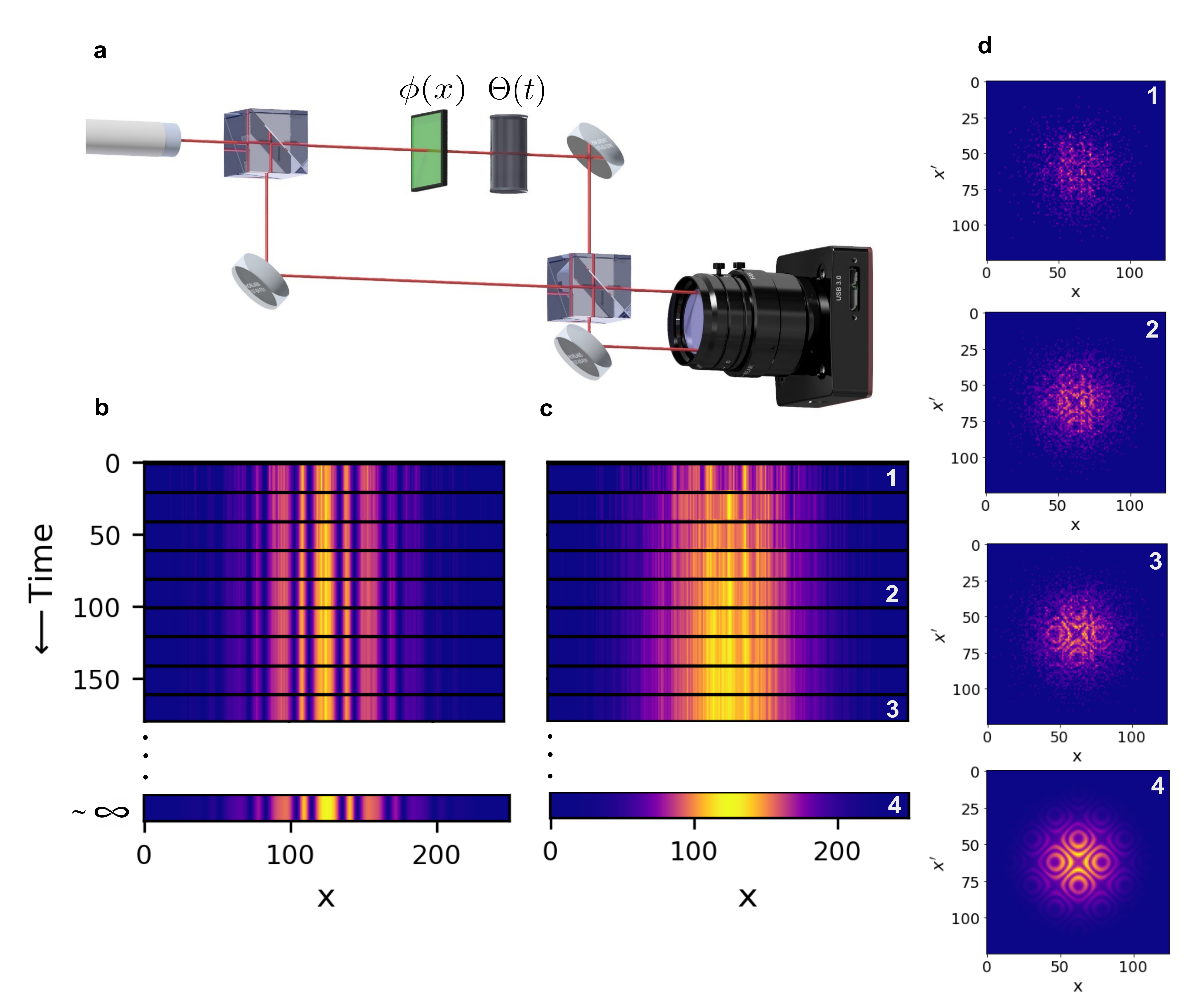}
   \caption[0.5\linewidth]{\textbf{(a)} Simplified schematic of the experiment: we divide input light into two paths, an object path, and a reference path. In the object path, we introduce a spatially varying 1D phase object $\phi(x)$ that we want to image. A time-fluctuating interferometric phase $\Theta(t)$ can be introduced to the system. For non-fluctuating phase $\Theta$, we can measure high visibility interference fringes \textbf{(b)}, see Eq.~\ref{eq1}. For rapidly fluctuating $\Theta(t)$, we observe the loss of interference visibility with increasing measurement time, whereas, for a short measurement time, the signal-to-noise ratio (SNR) is very low due to an insufficient number of detected photons \textbf{(c)}. In contrast, in our method, the visibility remains constant for an arbitrary acquisition time, which allows to create high-SNR interferograms \textbf{(d)}. Images \textbf{(b)} and \textbf{(c)} depict normalized one photon interference fringes for non-fluctuating and rapidly fluctuating phase scenarios respectively. Second-order correlation interferograms shown in \textbf{(d)} are created using the same photon detections as standard interferograms presented in \textbf{(c)}. Even for rapidly fluctuating phase, where we record only a few photons within the stability time of $\Theta(t)$, we can retrieve second-order correlation interferograms with high visibility and high SNR, that contain full information about the phase profile $\phi(x)$.}
    \label{fig:schem1}
\end{figure}
Since there is no information of $\phi(x,y)$ in this intensity pattern, the standard phase imaging schemes become inapplicable. One way to avoid the effect of time-dependent phase fluctuations is to shorten the duration of the measurement\cite{MAGYAR1963}. A short measurement time, however, reduces the amount of detected light and is therefore impractical in many cases including, for instance, imaging of photo-sensitive biological specimens, which require low-intensity light. Furthermore, for interferometric fluorescence super-resolution microscopy\cite{PhysRevLett.112.223602}, very low-intensity light{\cite{Morris2015}} often needs to be superposed. In such cases, any time-dependent phase fluctuations must be avoided due to the relatively long detection time requirement.
\par

Here, we introduce a method of phase imaging that is fully resistant to time-dependent phase fluctuations as long as it is possible to detect at least two photons within a time window in which the phase $\Theta$ is stable (phase stability time). The advantage of our method stems from the fact that it relies on the intensity correlation rather than intensity measurement, which makes it fundamentally different from standard phase imaging techniques\cite{OptHolography_book}. 
Numerous state-of-the-art approaches to low-light imaging \cite{Ndagano2022,Defienne2021,Lemos2014,Moreau2019,Zhang2021} rely on quantum light sources.
It is crucial, that our technique works for easily accessible and robust classical light sources, which dramatically broadens its applicability.
\par
The scheme of our experiment is illustrated in Fig.\,1. The object field is superposed with a reference field and the resulting interference pattern is detected by a camera. A time-dependent phase fluctuation $\Theta(t)$ is introduced to the reference field. Under these circumstances, no information on $\phi(x, y)$ can be retrieved from the intensity pattern given by Eq. (\ref{inten-incoh}), and therefore standard phase imaging techniques become inapplicable. 
Our phase imaging method is resistant to time-dependent phase fluctuations, provided that the fluctuating phase is spatially invariant throughout the entire sample \cite{Szuniewicz2019}. This approach is inspired by higher-order correlation effects observed in interference of light from independent sources \cite{Mandel1983}. It relies on measuring intensity correlations of light like in the intensity interferometry technique introduced by Hanbury Brown and Twiss (HBT)\cite{BROWN1956}. The HBT method and its generalizations were applied to a variety of light sources\cite{Hong1987,Ou1989,Pittman2003,Pfleegor1967,Rarity2005,Li2008,Bennett2009,Kim2013,Chrapkiewicz2016} and similarly, our technique might be applied in various scenarios including, for instance, laser and thermal light.

\par
We determine the correlation function between the intensities measured at a pair of points $(x,y)$ and $(x\prime,y\prime )$ (see Supplementary information S1 for detail), 
\begin{equation}
\left\langle \tilde{I}(x,y;t) \tilde{I}(x\prime ,y\prime;t) \right\rangle \propto 1\pm \frac{1}{2} \cos\left[\phi(x,y)-\phi(x\prime ,y\prime)\right],  
\label{inten-corr-incoh}
\end{equation}
where $\tilde{I}(x,y;t)$ is the instantaneous intensity measured at a point $(x,y)$ at time $t$. On the right-hand side of Eq. (\ref{inten-corr-incoh}), the plus ($+$) and minus ($-$) signs apply when the two points of measurement are in the same and different beam splitter outputs, respectively, we also assumed that $I_{\mathrm{r}}=I_{\mathrm{o}}$. Note that the information about the phase map of the object, which was lost in the intensity pattern [Eq. (\ref{inten-incoh}), Fig.\,1c], reappears in the intensity correlation map [Eq. (\ref{inten-corr-incoh}), Fig.\,1d]. 


\par
Remarkably, the 2nd-order intensity correlations map contains the full information required to optimally reconstruct $\phi(x,y)$ in the extreme case when only two photons are detected during the phase stability time. Our strategy of reconstructing the actual phase distribution is optimal in this scenario, which we prove rigorously using estimation theory tools, namely Fisher Information and Cram\'er-Rao bound \cite{cramer1999mathematical} (see Supplementary information S2 for detail).
\begin{figure}[h!]
   \centering
 \includegraphics[width=0.9\linewidth]{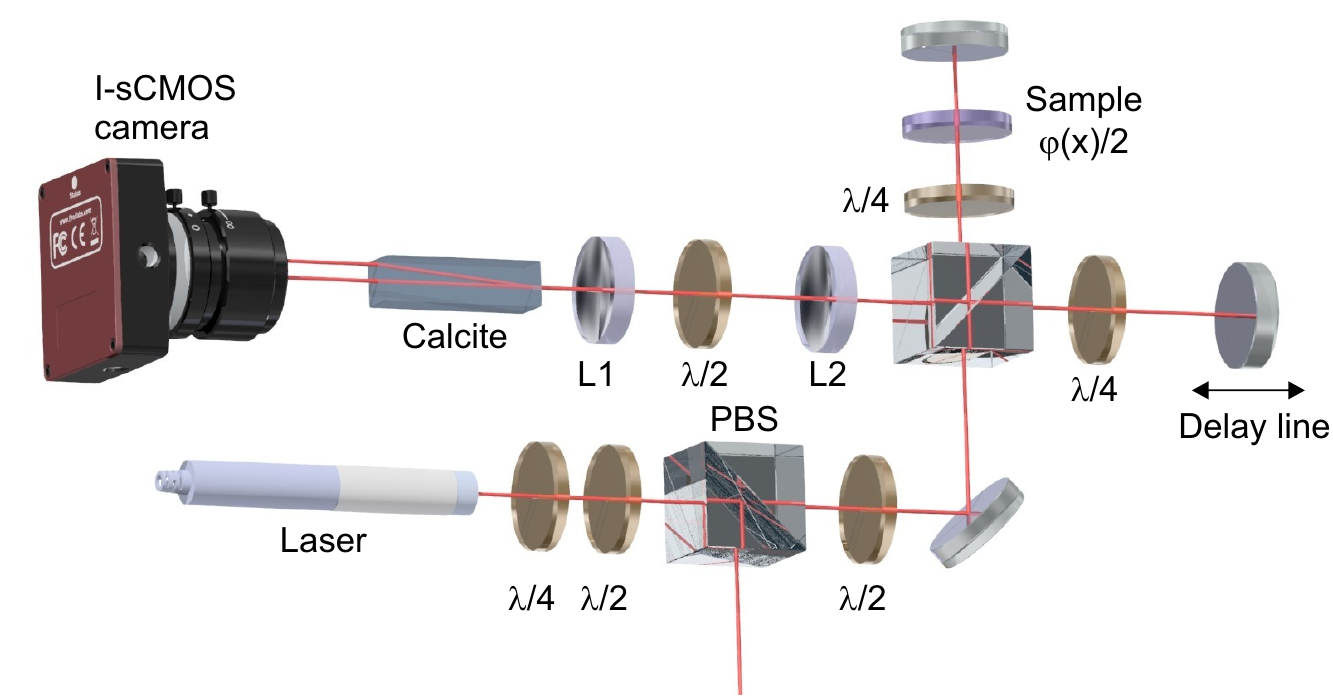}
   \caption[0.5\linewidth]{Experimental setup for noise-resistant phase imaging is constructed using polarization-based Michelson interferometer equipped with a 4f imaging system. Light from a 780 nm laser  passes through a half-wave plate ($\lambda/2$), quarter-wave plate ($\lambda/4$), polarization beam splitter (PBS) followed by additional $\lambda/2$---in this way, we control the beam intensity and polarization. Subsequently, the beam enters the interferometer. In one of its arms, the spatial phase $\phi(x)$ is imprinted onto the beam. The mirror in the other adds temporal fluctuations $\Theta(t)$ caused by a piezoelectric actuator. The beams from both arms are combined at the PBS, and the pass through a 4f imaging system consisting of two lenses (L1, L2), which are positioned such that the sample plane is imaged on the camera plane. The calcite polarizer combined with the output $\lambda/2$ acts as a 50/50 beam splitter. The intensified sCMOS (I-sCMOS) camera records single photons coming from both outputs of the interferometer. }

\label{fig:schem}
\end{figure}
\par
The idea demonstrated in Fig.\ref{fig:schem1} is implemented using a polarization-based Michelson interferometer equipped with an imaging system, shown in Fig.\, 2.  We perform experiments with three kinds of different phase masks $\phi(x)$ applied to the object beam. We imprint a 1D quadratic phase profile (Fig. 3a-b) to the beam by placing a cylindrical lens of a focal length, $f = 1000 \textrm{mm}$ in proximity to the mirror (Fig. 2). In a slightly modified setup (see Supplementary information S3), the sample is realized using a spatial light modulator (SLM), as it can display an arbitrary phase profile. The SLM imprints 1D sinusoidal (Fig.\, 4a-c) and exponential (Fig.\, 4d-f) phases to our object beam (see Supplementary information S3 for details).
A time-dependent fluctuating phase $\Theta(t)$ is introduced in the reference arm to make object and reference beams effectively incoherent. In order to register a very low photon flux and to minimize the exposure time, we use an Intensified-sCMOS camera.
To get a high interference visibility within a single frame, we keep the camera exposure time ($T_\mathrm{exp}$) low, such that fluctuations of $\Theta(t)$ are negligible within $T_\mathrm{exp}$. For $T_\mathrm{exp} \sim 20 \textrm{ns}$, this leads to a very low SNR for a single frame, as it contains only a few detected photons on average. However, this does not limit our method since we can sum many correlation maps created from different frames, even though the value of $\Theta(t)$ is different for each frame. As a consequence, we can perform quantitative phase imaging in the presence of temporal fluctuations of frequency of the order $\sim 1/T_\mathrm{exp} \simeq 50 \textrm{MHz}$. 
 This would be impossible for intensity-based interferometry---we would loose either visibility or SNR.
\par
\begin{figure}[h!]
   \centering
 \includegraphics[width=0.9\linewidth]{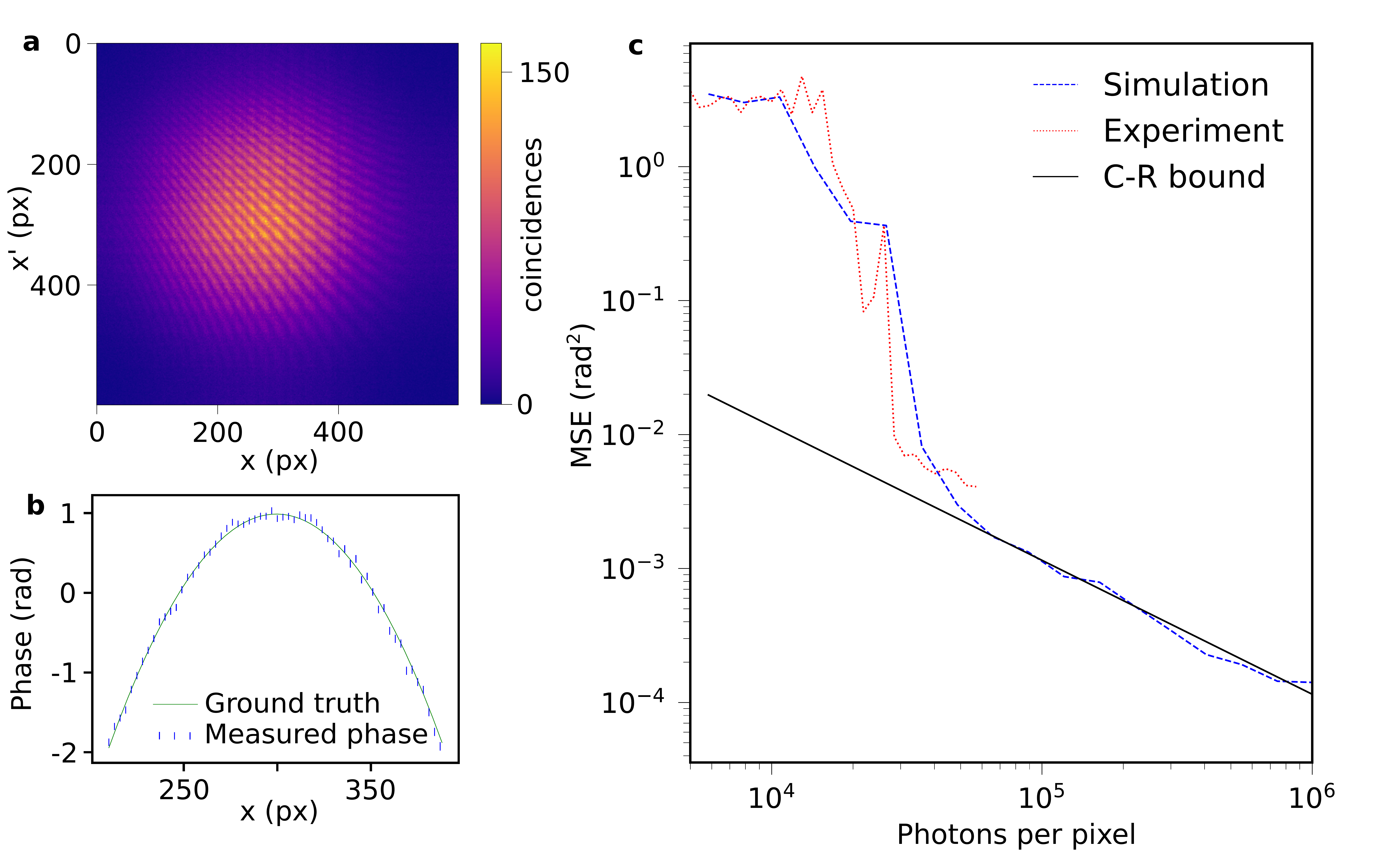}
   \caption[0.5\linewidth]{ Quantitative analysis of 1D quadratic phase profile reconstruction. \textbf{(a)} represents the experimentally measured coincidence map for a 1D quadratic phase profile, plotted with a solid line in \textbf{(b)}. The phase reconstructed from experimental data with error bars is also shown in \textbf{(b)}. The visibility of the fringes in the correlation map \textbf{(a)} is equal to $v^2/2$, where $v=0.6$ (theoretical maximum with classical light is $v=1$, which leads to $1/2$ visibility of the 2nd order interference). The total number of coincidences detected in the experiment is $ \sim 10^7$. By randomly removing a part of the collected signal, we can check how the MSE,  associated with the phase reconstruction, scales with the mean number of photons detected per pixel during the whole experiment $\textbf{(c)}$. The  MSE from the experiment is then compared with the MSE obtained using a simulated hologram, with the same parameters as in the experiment. We calculate the fundamental C-R lower bound on the MSE, assuming the visibility of hologram fringes to be equal to $0.6^2/2$ (as in our experiment). When no noise apart from shot noise is present (as in simulation), our method allows the saturation of this fundamental limit for a large enough ($\sim 5 \times 10^4$) number of photons detected per pixel. Other possible noise sources, e.g. camera dark counts, may slightly affect the MSE obtained experimentally. }
\label{fig:schem}
\end{figure}
\par
\begin{figure}[htp]
   \centering
 \includegraphics[width=0.85\linewidth]{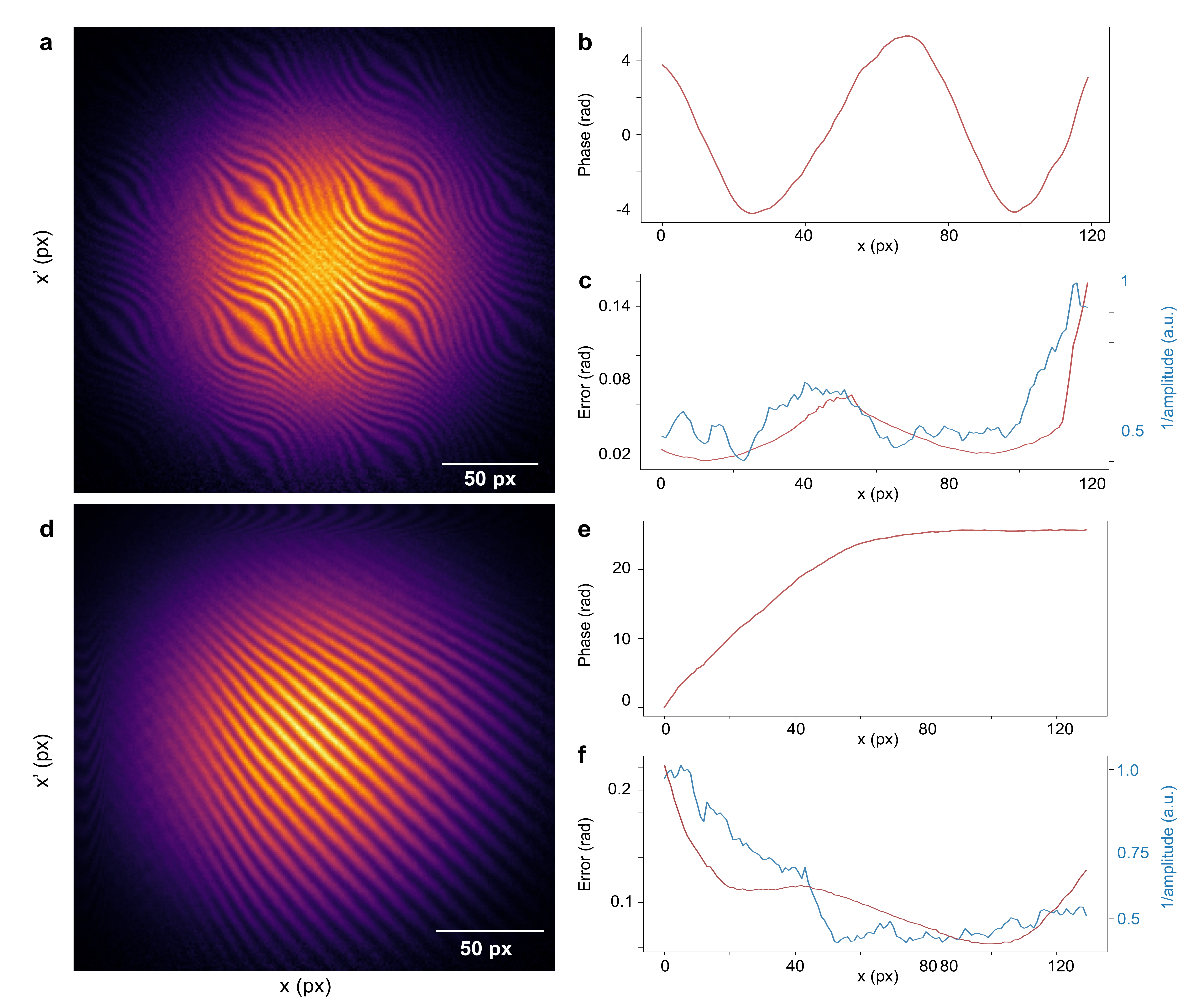}
   \caption[0.5\linewidth]{Experimental results of the spatial phase profiles displayed by SLM. Measured coincidence maps (correlation functions) between outputs of the interferometer for  sinusoidal, and  exponential phases are shown in \textbf{(a)} and \textbf{(d)} respectively. 
   \textbf{(b)} and \textbf{(e)} represent the  aforementioned reconstructed phases. \textbf{(c)} and \textbf{(f)} show errors and inverse square-root of the intensities.} 
\label{fig:schem}
\end{figure}
\section*{Results}
In our experiment, each frame contains on average $ \sim 15$ photons at both outputs of the interferometer per frame. We remove temporal correlations between subsequent frames by randomly permuting the order of frames before further processing---this does not change the performance of our method but allows us to simulate the conditions, in which the phase $\Theta(t)$ is independently sampled for each consecutive frame. Such conditions occur, when the noise frequency is larger than the delay between subsequent exposures. It is then impossible to retrieve phases using standard interferometric methods. Averaging recorded intensities over multiple subsequent frames would result in a loss of the visibility of the interference fringes. In contrast, we average correlations of detected photons' positions without any loss of the phase information. Such averaging over multiple frames results in the reproduction of the correlation function (Eq.3), from which we can retrieve the phase profile using standard digital holography methods---Fourier off-axis holography\cite{Mertz} has been used in our case. This analyzing mechanism is the essence of our noise-resistant phase imaging technique. See the supplementary information S4 for details.

To quantitatively assess the precision of our phase reconstruction method, we measure a 1D quadratic phase object and compare the mean squared error (MSE) obtained in the experiment and in a simulation with the fundamental lower-bound for the MSE given by the inverse of Fisher Information, namely Cram\'er-Rao bound (C-R bound), see Supplementary information S2 for details.
The measured coincidence map (Fig.\,3(a)) consists of approximately $10^7$ registered photon pairs. To simulate the most extreme case, we use only two randomly chosen photons from each frame for further processing. We estimate the phase profile shape using the collected data, and compute the associated MSE, which is the average square distance between the real (ground truth) and measured phase (Fig.\,3(b)). As we show in Fig.\,3(c), the MSE drops down with the total number of measured photons, and eventually reaches the theoretical minimum. This proves, that our method of phase estimation is optimal when at most two photons are measured during the phase stability time---notice, that this is the most extreme limit in which one can gain any information about the phase profile. 
\par
To show the versatility of our method, we also reconstruct SLM-encoded phase profiles $\phi(x)$ of different shapes---sinusoidal and exponential. Figures 4(a/e), (b/f), and (c/g) represent the measured hologram, the retrieved phase, and the error per pixel for the sinusoidal/exponential phase $\phi(x)$ Extra systematic errors of phase recovery in both cases are caused by the SLM imperfections.
\par
\section*{Discussion}
We have demonstrated a complete retrieval of phase patterns in the presence of time-dependent high-frequency phase fluctuations using spatial intensity correlations measurement. Since this time-dependent phase fluctuates much faster compared to the detection time, the reference and the object fields can be considered mutually incoherent \cite{MW}. Consequently, our method is applicable to  the case when reference and object fields are generated from independent sources \cite{Mandel1983,RevModPhys.58.209,Rarity2005,Ou1988}. Although we performed the experiment with laser light, our method is applicable to optical fields characterized by different statistics, such as thermal light. What is more, our technique provides a quantitative phase image and an amplitude image simultaneously. The latter can be easily accessed by summing the intensity correlation interferogram over one of the dimensions.

We stress that the presented method optimality is proven using the C-R bound\cite{cramer1999mathematical}---all the spatial phase information contained in the detected photons is retrieved with our technique. 

High temporal resolution (short exposure time $T_\mathrm{exp}$) is necessary for overcoming the problem of the rapidly fluctuating temporal phases.
In our experiment, we used I-sCMOS camera, which allows to achieve $T_\mathrm{exp} \sim 20 \textrm{ns}$, corresponding to $50 \textrm{MHz}$ maximal phase noise frequency we can deal with.
However, our method is not limited to this type of camera and can be implemented using various high-temporal resolution detection platforms. Because of high quantum efficiency, high temporal resolution, and low noise level in recent single-photon avalanche diode (SPAD) array technology\cite{Antolovic2018}, our method can also be implemented by SPAD arrays. The technique works both in the photon counting regime and by employing less accurate intensity measurements, yet it is the most remarkable for cases where registering more than two photons per phase stability time is rare. Our method can be readily generalized to two-dimensional spatial phase profiles by creating higher-dimensional correlation maps. It also allows for implementation in different degrees of freedom, such as temporal or spectral, allowing the creation of joint probability maps both for photon detection times or their detected wavelengths. It is also possible to incorporate an additional degree of freedom to a measurement, measuring for instance joint temporal-spatial correlation maps.

Equation 3 is only exact when all the values of $\Theta$ have the same probability of appearing during the time interval in which the whole measurement is performed. To satisfy this condition for arbitrary temporal phase noise, it is enough to add random uniformly distributed signal oscillating between $0$ and $2 \pi$ to the unknown  phase fluctuations $\Theta(t)$. 
\par

Our method opens up possible applications in quantitative phase imaging under low light conditions for microscopy as well as for fundamental research.
Unbalanced interferometers, such as the ones used in the time-bin encoding, could be of particular interest, as our method enables the use of additional degrees of freedom (for multi-dimensional information encoding) while filtering out phase fluctuations arising from unmatched optical paths.

Moreover,  X-rays and matter waves\cite{Zanette2010,Weitkamp2005}, because of their shorter wavelengths,  require much tighter alignment and better mechanical stability to interfere. Since our technique is phase noise resistant, it holds potential for phase-sensitive imaging using X-ray and matter waves \cite{Rasel1995,Arndt2012}. 

It is of particular interest for future work to investigate the possibility of in-situ correlation measurement using non-linear effects, such as second harmonic generation or two-photon absorption. It could be realized by creating a setup superimposing an original fringe pattern and one rotated by 90 degrees. A nonlinear readout can act as a coincidence detector\cite{boitier2009} and therefore allow to average on the camera chip, fulfilling the requirement for fast measurement by the instantaneous coincidence detection - therefore allowing an almost unlimited noise bandwidth rejection.

\par

\section*{Acknowledgments}
We acknowledge discussions with Piotr Węgrzyn, Łukasz Zinkiewicz, Michał Jachura, Wojciech \\Wasilewski, Adam Widomski, Michał Karpiński, Mariusz Semczuk, and Marek Żukowski. This work was supported by the Foundation for Polish Science under the FIRST TEAM project 'Spatiotemporal photon correlation measurements for quantum metrology and super-resolution microscopy' co-financed by the European Union under the European Regional Development Fund (POIR.04.04.00-00-3004/17-00). JS acknowledges that this research was funded in part by the National Science Centre, Poland, Grant number: 2022/45/N/ST2/04249.

\section*{Supplementary informations}
\par

\textbf{S1} -  Statistical model of the experiment
\\
\\
\textbf{S2} - Fundamental precision limits in interferometric phase imaging
\\
\\
\textbf{S3} - Experimental setup details
\\
\\
\textbf{S4} - Data analysis

\bibliography{scibib}

\bibliographystyle{Science.bst}

\clearpage

\end{document}


\title{Supplementary information and methods}
\date{}
\maketitle

\section*{S1: Statistical model of the experiment}
 Two cameras are set on the two outputs of the interferometer, and each of them consists of the same number of pixels ($n_{\mathrm{pix}}$). The sample area with phase $\phi_i$ is imaged onto the pixel number $i$ on both cameras. Only two photons are received per the stability time of the interferometer phase. A single measurement consists of the detection of these two photons. The output of the single measurement is a pair $(i_{+/-}, j_{+/-})$. Numbers $i,j$ stand for the numbers of pixels in which photons were detected, whereas indices $+$ or $-$ indicate in which of the two outputs the corresponding photon was measured. The probability of measuring a single photon in a pixel $i_{+/-}$ is:
\begin{equation}
\label{single}
p(i_{+/-}, \Theta) = \tilde{\mathcal{N}} I_i \frac{1}{2} (1 \pm v \cos(\phi_i + \Theta)),
\end{equation}
where $ \tilde{\mathcal{N}}$ is a normalization factor, $v$ is interferometer visibility,  $\Theta$ is an extra, spatially uniform, possibly fluctuating phase, and $I_i$ is the intensity of the beam illuminating the phase mask in the area corresponding to pixel $i$. Phase $\Theta$ is stable within the time of detection of a single photon pair, its value for each pair is independently drawn from the continuous uniform probability distribution $p(\phi) = \frac{1}{2 \pi}$ for $\phi \in [0, 2\pi]$. We do not have access to the randomly chosen value  of $\Theta$, so the observed probability of measuring pair $(i_{+/-}, j_{+/-})$ in every single frame is:
\begin{equation} \label{joint}
p(i_{+/-}, j_{+/-}) = \int _0 ^{2 \pi}  p(i_{+/-}, j_{+/-}, \Theta) d \Theta,
\end{equation}
 where $p(i_{+/-}, j_{+/-}, \Theta) = p(i_{+/-}, \Theta) p(j_{+/-}, \Theta) $ is a joint probability distribution of measuring pair $(i_{+/-}, j_{+/-})$ with the fixed value of $\Theta$.
From Eq. \ref{joint}, we obtain the formulas:
\begin{equation} \label{same}
p(i_+, j_+) = p(i_-, j_-) =  2 \mathcal{N} I_i I_j \left(1+\frac{v^2}{2} \cos (\phi_i-\phi_j) \right)
\end{equation}
\begin{equation} \label{diff}
 p(i_+, j_-) = p(i_-, j_+) = 2 \mathcal{N} I_i I_j \left(1-\frac{v^2}{2} \cos (\phi_i-\phi_j) \right),
\end{equation}
where $\mathcal{N}$ is a new normalization factor. Notice, that Eq. (3) from the main text is retrieved from the two above equations after substituting $v=1$ and neglecting the normalization factor. The above equations are the starting point for further inference about the maximal precision of the measurement. Full information about every single measurement is included in the dependence of the probability $p(i_{\pm}, j_{\pm})$ on the estimated parameters $\phi_i$.

\section*{S2: Fundamental precision limits in interferometric phase imaging}
\subsection*{S2.1 Rapid fluctuations regime}
In order to calculate the maximal precision of estimation of the parameters $\phi_i$, Fisher Information (FI) matrix $F$ will be calculated. Then the inverse of the covariance matrix for all sets of unbiased estimators $\tilde \phi_i$ is lower bounded by the inverse of FI matrix \cite{cramer1999mathematical}. There are 4 different types of events, which can occur during one experiment - two photons may be detected in one output ($+$ or $-$) or in different outputs ( we distinguish between $+-$ and $-+$). We can distinguish between these 4 types, so the FI is the sum of FI matrices for all events types:
\begin{equation}
F = F^{++} + F^{--} + F^{+-} + F^{-+}
\end{equation}
From equations \ref{same} and \ref{diff} we can simply conclude, that $F^{++} = F^{--}$ and $F^{+-} = F^{-+}$. Let us now focus on the calculation of  $F^{++}$ matrix. \newline \newline
In order to simplify the formulas, the following notation will be used:
$$
p(i_{+}, j_{+}) \equiv p(i,j), ~~ \frac{\partial}{\partial \phi _k} \equiv \partial_k
$$
The elements of the $F^{++}$ can be written in the following form:
\begin{equation} \label{fisher}
F^{++}_{kl} = \sum _{i,j=1} ^{n_{\mathrm{pix}}} \frac{\partial _k p(i,j) \partial _l p(i,j)}{p(i,j)},
\end{equation}
Moreover,
\begin{equation}
\partial _k p(i,j) = \mathcal{N} I_i I_j v^2 (\delta_{jk} - \delta_{ik}) \sin (\phi_i - \phi_j),
\end{equation}
\begin{equation}
\partial _k p(i,j) \partial _l p(i,j) = (\delta_{jk} - \delta_{ik}) (\delta_{jl} - \delta_{il}) \mathcal{N}^2 I_i^2 I_j^2 v^4 \sin^2 (\phi_i - \phi_j),
\end{equation}
where $\delta_{ij}$ is a Kronecker delta.
Consequently,
\begin{equation}
F^{++}_{kl} = \sum _{i,j=1} ^{n_{\mathrm{pix}}} \frac{(\delta_{jk} - \delta_{ik}) (\delta_{jl} - \delta_{il}) \mathcal{N} I_i I_j v^4 \sin^2 (\phi_i - \phi_j) }{2+v^2 \cos(\phi_i - \phi_j)}
\end{equation}
If $k \neq l$, then for any $m$ we have $ \delta_{mk} \delta_{ml} = 0$, so $(\delta_{jk} - \delta_{ik}) (\delta_{jl} - \delta_{il}) = - \delta_{jk} \delta_{il} - \delta_{ik} \delta_{jl}$.
That means, that non-diagonal matrix elements are:
\begin{equation}
F^{++}_{kl} = - \frac{2 \mathcal{N} I_k I_l v^4 \sin^2 (\phi_k - \phi_l) }{2+v^2 \cos(\phi_k - \phi_l)} ~ , ~ ~ k \neq l
\end{equation}
With the help of the equality $(\delta_{jk} - \delta_{ik})^2 = \delta_{jk} + \delta_{ik} - 2 \delta_{ik} \delta_{jk}$ we can obtain diagonal terms of $F^{++}$:
\begin{equation}
F^{++}_{kk} = 2 \mathcal{N} I_k v^4 \sum_{i=1}^{n_{\mathrm{pix}}} \frac{ I_i \sin^2 (\phi_i - \phi_k)}{2+v^2 \cos(\phi_i - \phi_k)} \label{fkk}
\end{equation}
For any function $f$:
\begin{equation}
\sum_{i=1}^{n_{\mathrm{pix}}} f( \phi_i, I_i) = n_{\mathrm{pix}} \langle f(\phi_i, I_i)\rangle_i, 
\end{equation}
where $ \langle f(\phi_i, I_i) \rangle_i$ is the mean value of the function over all pixels. From now on, we assume that the number of pixels is big and that each phase in the sample occurs with the same frequency. What is more, the intensity of illuminating beam $I_i$ is assumed to change slowly compared to the change of phase $\phi_i$. In other words, many different phases occur in the region with approximately constant intensity. From these assumptions, we obtain the equality
\begin{equation}
\langle f(\phi_i, I_i) \rangle_i = \frac{1}{2 \pi} \int_0 ^{2 \pi} f(\phi, \langle I \rangle) d \phi , 
\end{equation}
  which is true when $f$ is linear with its second argument $I$,  $\langle I \rangle$ stands for the mean intensity of the illuminating beam. Using the above assumptions, we can rewrite equation \ref{fkk} as:
\begin{equation}
F_{kk}^{++} = 2 \mathcal{N} I_k \langle I \rangle v^4 \frac{n_{\mathrm{pix}}}{2 \pi} \int_0^{2 \pi} \frac{ \sin^2 (\phi - \phi_k)}{2+v^2 \cos(\phi - \phi_k)} d \phi
\end{equation}
Consequently, diagonal terms of $F^{++}$ are:
\begin{equation}
F_{kk}^{++} = 2 \mathcal{N} \langle I \rangle I_k n_{\mathrm{pix}} ( 2- \sqrt{4-v^4})
\end{equation}
Let us now calculate the normalization factor $\mathcal{N}$. We will use the fact, that sum of probabilities of all events must be equal to one:
\begin{equation}
\sum_{i,j = 1}^{n_{\mathrm{pix}}} p(i_+, j_+) + p(i_+, j_-) + p(i_-, j_+) + p(i_-, j_-) = 1
\end{equation}
Using equations \ref{same} and \ref{diff} we obtain:
\begin{equation}
8 \mathcal{N} \sum_{i,j=1}^{n_{\mathrm{pix}}} I_i I_j =1
\end{equation}
We can rewrite the sum in the above equation as:
\begin{equation}
\sum_{i,j=1}^{n_{\mathrm{pix}}} I_i I_j =  \left( \sum_{i=1}^{n_{\mathrm{pix}} }I_i \right) ^2 = n_{\mathrm{pix}}^2 \langle I \rangle ^2
\end{equation}
and obtain:
\begin{equation}
\mathcal{N} = \frac{1}{8 n_{\mathrm{pix}}^2 \langle I \rangle ^2}
\end{equation}
Finally, $^{++}$ matrix can be written in the form:
\begin{equation}
F^{++}_{kl} = \left\{ \begin{array}{ll}
\frac{1}{4 n_{\mathrm{pix}}} \frac{I_k}{\langle I \rangle} ( 2 - \sqrt{4-v^4}) & \textrm{for $k = l$}\\ \\
- \frac{1}{4 n_{\mathrm{pix}}^2} \frac{I_k I_l}{\langle I \rangle ^2} \frac{2  v^4 \sin^2 (\phi_k - \phi_l) }{2+v^2 \cos(\phi_k - \phi_l)} & \textrm{for $k \neq l$}\\

\end{array} \right.
\end{equation}
We have calculated the $F^{++}$ matrix, which is obviously the same as $F^{--}$ matrix, because formulas for probabilities in both cases are the same. Analogous  calculation show, that also $F^{+-} = F^{-+} = F^{++}$. Using the FI additivity we obtain the terms of $F$ matrix:
\begin{equation}
F = 4 F^{++}
\end{equation}
This is the FI matrix associated with the measurement of a single frame. If the total number of $n_\mathrm{phot}$ photons is detected in the experiment (which means $n_\mathrm{phot}/2$ independent photon pairs), then, from the Cram\'er-Rao bound, 



\begin{equation}
 \Delta ^2 \phi_k \ge \frac{1}{n_\mathrm{phot}/2}  (F^{-1})_{kk}
\end{equation}
In general, the estimator which satisfies the above inequality may not exist, however, it is possible to get arbitrarily close to the above bound if the number of measurements is big enough. That means, that the inequality becomes an equality if $ n_\mathrm{phot} \to \infty$.
To simplify the calculations we also use the inequality:
\begin{equation}
(F^{-1})_{kk} \ge (F_{kk})^{-1},
\end{equation}
which is true for all hermitian $F$. The above inequality is not saturated in general, especially when non-diagonal terms of $F$ are significant. However, in our case, the non-diagonal terms are asymptotically $n_{\mathrm{pix}}$ times smaller than diagonal terms. $n_{\mathrm{pix}}$ is also size of the $F$ matrix. It may be proven, that for such scaling of non-diagonal terms with the size of the matrix, the above inequality becomes saturable for $n_{\mathrm{pix}} \to \infty$. Using both of the above inequalities, we obtain the following bound:
\begin{equation}
\Delta \phi _k \ge \sqrt{\frac{n_{\mathrm{pix}} \langle I \rangle }{n_{\mathrm{phot}} I_k}} \frac{1}{\sqrt{1-\sqrt{1-v^4/4}}}
\end{equation}
The value $ n_k = \frac{n_{\mathrm{phot}} I_k}{n_{\mathrm{pix}} \langle I \rangle }$ may be interpreted as the expected number of photons detected in pixel number $k$. The above bound may be rewritten in the intuitive form:
\begin{equation} \label{pair}
\Delta \phi_k \ge \sqrt{\frac{1 }{n_k}} \frac{1}{\sqrt{1-\sqrt{1-v^4/4}}}
\end{equation}
From this form of inequality, it's clear, that the accuracy of measuring the value of the particular phase depends directly on the number of photons interacting with the measured area.
\subsection*{S2.2 Comparison with slow fluctuations regime}
Let's compare our result with the phase estimation precision limit for an interferometer with a slowly fluctuating phase $\Theta$. First of all, let's notice that we can't beat the accuracy achievable in the situation, in which extra phase $\Theta$ is known for all the detected photons---the information we get in a situation with unknown $\Theta$ is always smaller, even if the stability time of the interferometer is bigger. If $\Theta$ values are known, each single photon detection could be treated as an independent event (which was not the case in the previous section). Let's calculate the FI 
 matrix for the single photon detection when $\Theta$ is fixed. A single measurement is fully described by the probability distribution from equation \ref{single}, and
\begin{equation} \label{part}
 \partial_k p(i_{+/-}) =  \mp \frac{1}{2} \delta_{ki} \tilde{\mathcal{N}} I_i  v \sin ( \phi_i + \Theta).
\end{equation}
In the case with fixed $\Theta$, the one-photon FI matrix is
\begin{equation}
F^{(1)}_{kl} = \sum _{i=1} ^{n_{\mathrm{pix}}} \frac{\partial _k p(i_{+}) \partial _l p(i_{+})}{p(i_{+})} + \sum _{i=1} ^{n_{\mathrm{pix}}} \frac{\partial _k p(i_{-}) \partial _l p(i_{-})}{p(i_{-})}.
\end{equation}
From equation \ref{part} it is clear, that all non-diagonal terms of the $F^{(1)}$ matrix are equal to zero. This is because we obtain information about the $\phi_i$ phase only in case of detection of a photon in the pixel $i_{+/-}$. The diagonal terms are
\begin{equation} \label{30}
F^{(1)}_{kk} =  \tilde{\mathcal{N}} I_i \frac{ v^2 \sin^2(\phi_i + \Theta)}{1 - v^2 \cos^2(\phi_i + \Theta)}.
\end{equation}
To make this case similar to the case described in the previous section let's assume, that $\Theta$ fluctuates and each value of $\Theta$ appears with the same frequency ( the difference is that $\Theta$ fluctuates slowly and we know its value). Then the mean FI for the single measurement is:
\begin{equation}
 \langle F^{(1)}_{kk} \rangle _\Theta = \frac{1}{2 \pi} \int_0^{2 \pi} F^{(1)}_{kk} d \Theta =  \frac{I_i}{n_{\mathrm{pix}} \langle I \rangle} \left( 1- \sqrt{1-v^2} \right) , 
\end{equation}
where formula $\tilde{\mathcal{N}} = \frac{1}{n_{\mathrm{pix}} \langle I \rangle}$ obtained from the normalization condition was used. If we define $n_k = \frac{n_{\mathrm{phot}} I_k}{n_{\mathrm{pix}} \langle I \rangle }$ as in the previous section, we obtain the best possible accuracy of measuring each phase $\phi _k$:
\begin{equation} \label{many}
\Delta \phi _k \ge \frac{1}{n_\mathrm{phot} \sqrt{F_{kk}^{(1)}}} = \sqrt{\frac{1 }{n_k}} \frac{1}{\sqrt{1-\sqrt{1-v^2}}}
\end{equation}
Equation \ref{many} is very similar to the equation \ref{pair}- the only difference is that term $\frac{v^4}{4}$ is substituted by the term $v^2$. That means, that the fact that one has only two photons per phase fluctuations stability time, leads to a decrease of the effective visibility of the interferometer from $v$ to $\frac{v^2}{2}$, compared to a slowly fluctuating case, in which we can assume, that we know the value of $\Theta$. 
\section*{S3: Experimental setup details}
Our experimental setup comprises a polarization-based Michelson interferometer equipped with a $4f$ imaging system. As a light source, we use a diode laser at a wavelength of $780 nm$ coupled to a single-mode fiber. At the output of the fiber, for intensity and polarization control, the beam passes through a half-wave plate, a quarter-wave plate, and polarizing beam splitter (PBS), and another half-wave plate, and then enters the interferometer. Each of the two paths in the interferometer is encoded with orthogonal polarization. In order to imprint different kinds of phase profiles $\phi(x)$ on the object beam, we build two kinds of slightly modified setups---one with a cylindrical lens placed in front of the mirror in the horizontally polarized light beam path in the interferometer, while in the other setup, we replace the mirror in the same path with a spatial light modulator (SLM). The interferometric mirror in the reference arm is given a phase fluctuation by attaching it to a piezoelectric actuator driven by a ramp function signal.
\begin{figure}[h!]
   \centering
 \includegraphics[width=0.9\linewidth]{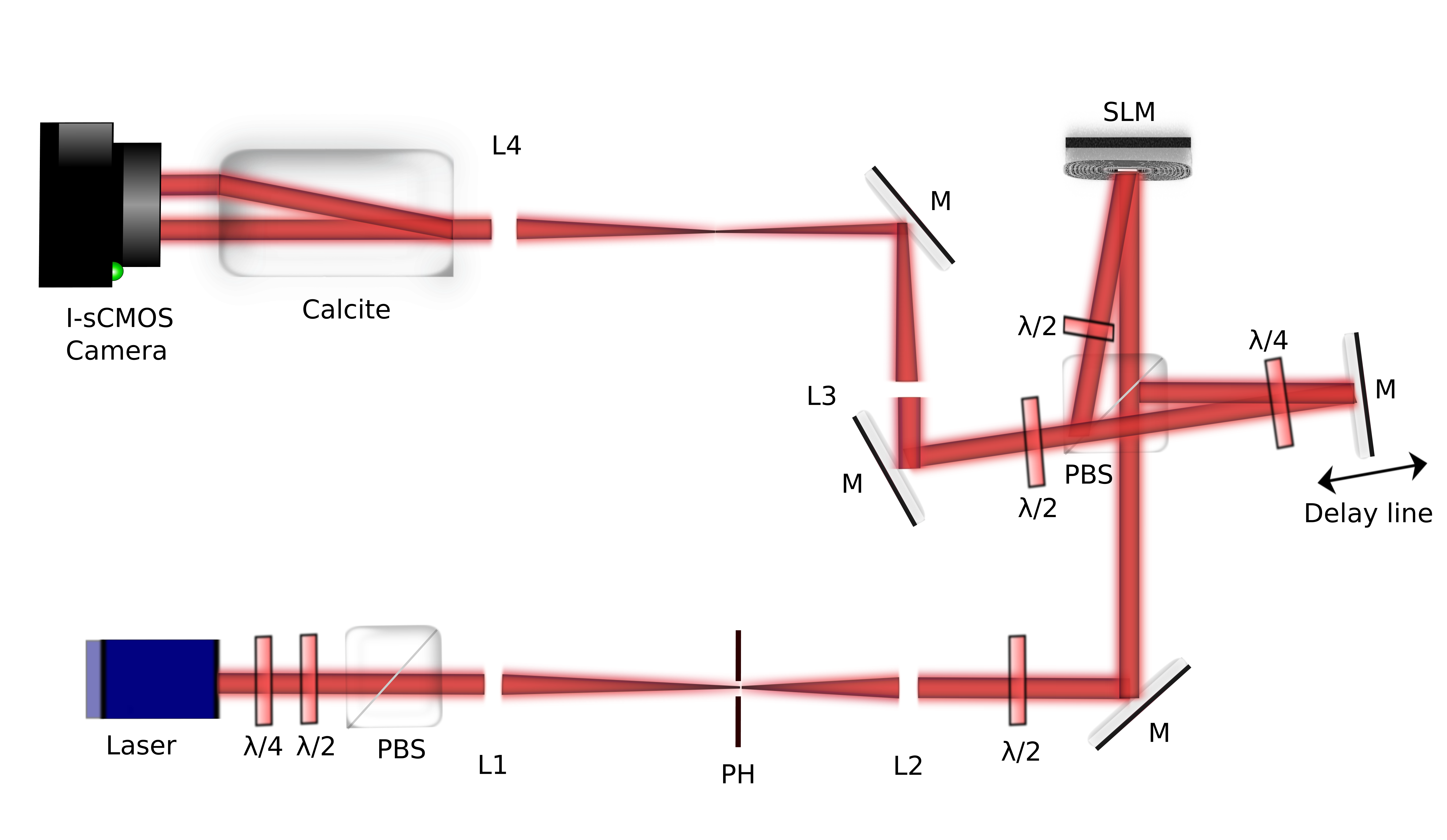}
   \caption[0.5\linewidth]{Experimental setup for noise-resistant phase imaging with SLM.}
\label{fig:schem}
\end{figure}

We perform experiments with three kinds of different phase masks applied to our object beam. Our first configuration is to imprint a one-dimensional quadratic local phase profile to the beam by placing a cylindrical lens of focal length, $f = 1000~ \textrm{mm}$ in proximity to the mirror (Fig. 2 in the main text). Additionally, in our second configuration with SLM (from the HOLOEYE PLUTO) as a phase mask, we can display any arbitrary phase profile. As an example, we imprint one-dimensional exponential and sinusoidal phases onto our object beam. Since the SLM is efficient for only one polarization of light (in our case horizontal), we place the SLM in the horizontal path of the interferometer and the reflected beam with a given phase mask (the object beam) passes through a $\lambda/2$ plate. We introduce a time-dependent phase fluctuation $\Theta(t)$ in the reference arm - vertically polarized beam path in the interferometer) to make it effectively incoherent with the object beam. This is realized with a piezoelectric actuator driven by a ramp of 1.234 Hz. This shouldn't be confused with the maximal noise frequency for which our method works. Both the object and reference beams are combined on the polarizing beam splitter (PBS). At the output of the interferometer, we have a 4f imaging system consisting of lenses L3 and L4 of focal length 200mm each to image the SLM plane (phase plane) the intensified sCMOS (I-sCMOS - with the image intensifier from Hamamatsu V7090D-71-G272 and sCMOS from Andor Zyla) camera. To observe the interference, the orthogonally polarized object and the reference beam are required to be indistinguishable, and to do so, we rotate the polarization of both beams by 45 degrees with a half-wave plate and we perform the projective measurement in the original bases with a calcite crystal. Here, the calcite along with the waveplate act as a 50/50 Beamsplitter. This mixes the light from both outputs and allows us to observe interference in both outputs of the splitter. The I-sCMOS camera records single photons at both outputs of the interferometer on two regions.
In order to get a high interference visibility within a single frame, we keep the camera exposure time low such that fluctuations are negligible within one camera frame. The use of short exposure time of the I-sCMOS, in the single nanosecond timescale, gives it stability and resistance against fluctuations up to tens of MHz. The interference visibility is slightly reduced due to imperfect imaging because of the path length difference in the calcite.
We collect the data with a 200 Hz of frame rate.

\section*{S4: Data analysis}

In this section, we describe how to obtain the final phase profiles from the raw experimental data. We first describe a standard approach for a non-fluctuating phase $\Theta$, which we treat as verification of the ground truth. Then, we will follow with the analysis of the phase-fluctuating case and quantitative analysis of the reconstruction precision.

\subsection*{S4.1 One dimensional non-fluctuating phase}
For ground truth measurements, we stabilize the interferometer with a box to reduce the airflow in the setup and use a laser beam to record bright interference fringes in a single shot. We record these fringes using a standard CMOS camera (Matrix Vision mvBlueFOX-IGC). We are then blocking interferometer paths, one at a time, to record beam profiles. We use these profiles to normalize the interference fringes and get an interferogram not modulated by the input beam profile.

We add padding---increase the size of the image from $n_{\mathrm{pix}}$ by $n_{\mathrm{pix}}$ pixels to $3 n_{\mathrm{pix}}$ by $3 n_{\mathrm{pix}}$ pixels by adding an $n_{\mathrm{pix}}$ pixel ramp function on each side of the array from the image's edge value to zero (see NumPy library documentation for full description: numpy.pad, mode = 'linear\_\_ramp', end\_values = 0). Adding padding increases the available Fourier space and allows us to get a smoother phase retrieval. Then the normalized and padded interferogram is converted to the Fourier space using a fast Fourier transform (FFT). Then we extract half of the data, removing low frequencies in the middle. This is a common practice for phase retrieval~\cite{mertz_2019} and allows us to directly retrieve the phase---by inversing the FFT we get the phase as an argument of complex numbers from the resulting map. 

The reconstructed phase is still wrapped---an argument lies in the range from 0 to $2\pi$ so the phase needs to be unwrapped; we perform this using the SciPy library in Python. At the end of the process, we get a two-dimensional phase profile, and since the analyzed phase is one-dimensional, we average the retrieved phase within the area of sufficient signal strength over one of the dimensions. We end up with a one-dimensional phase profile that was created on the basis of standard, well-established interferometric method and that can be treated as the basis for further experiments.

\subsection*{S4.2 One Dimensional Fluctuating Phase}
\label{Supp:analysis_fluct}
For the creation of the joint probability map, we record the positions of all photons. We are considering only one-dimensional cases, so it is enough to know the position along the direction in which the phase changes. We extract the correct positions and create a square array with dimensions of $n_{\mathrm{pix}}$ by $n_{\mathrm{pix}}$, where $n_{\mathrm{pix}}$ is the number of pixels rows for the single-photon camera. This square array is our correlation map---its element with coordinates $(i,j)$ contains the number of frames in which a pair of photons, one at position $i$, another at position $j$, was detected. In principle, two photons detected per frame are enough to create such a correlation map, given a sufficient number of measured frames.

We create correlation maps for all frames and sum them. We finally get an average joint probability map that contains phase information $\phi(x)$:
\begin{equation}
\left\langle \tilde{I}(x;t) \tilde{I}(x\prime ;t) \right\rangle \propto 1\pm \frac{v^2}{2} \cos\left[\phi(x)-\phi(x\prime)\right],  
\label{inten-corr-incoh}
\end{equation}
where $v^2/2$ is the visibility of 2nd order intereferogram fringes. We use notation $v^2/2$ because in an ideal scenario, visibility of 1st order fringes $v$ (observed for non-fluctuating case) would lead to visibility $v^2/2$ observed for fluctuating case for a 2nd order interference. Notice, that the above equation is equivalent to Eq. (3) from the main text---the difference is that now we consider 1D phase profile and imperfect interference visibility.
Please note that at this point we can use exactly the same phase retrieval method as for the non-fluctuating case, as the phase information is imprinted in a form resembling the first-order interference: $\cos\left[\phi(x,y)-\phi(x\prime ,y\prime)\right]$. We use the same data analysis as in Section 3.1 of this document. The only difference is that we normalize the joint intensity map using a sum of all interferograms, instead of the sum of separate measurements, as the phase information in the first-order interferogram is lost (please refer to the normalization steps in Fig. 2).

\begin{figure}[h!]
   \centering
 \includegraphics[width=0.9\linewidth]{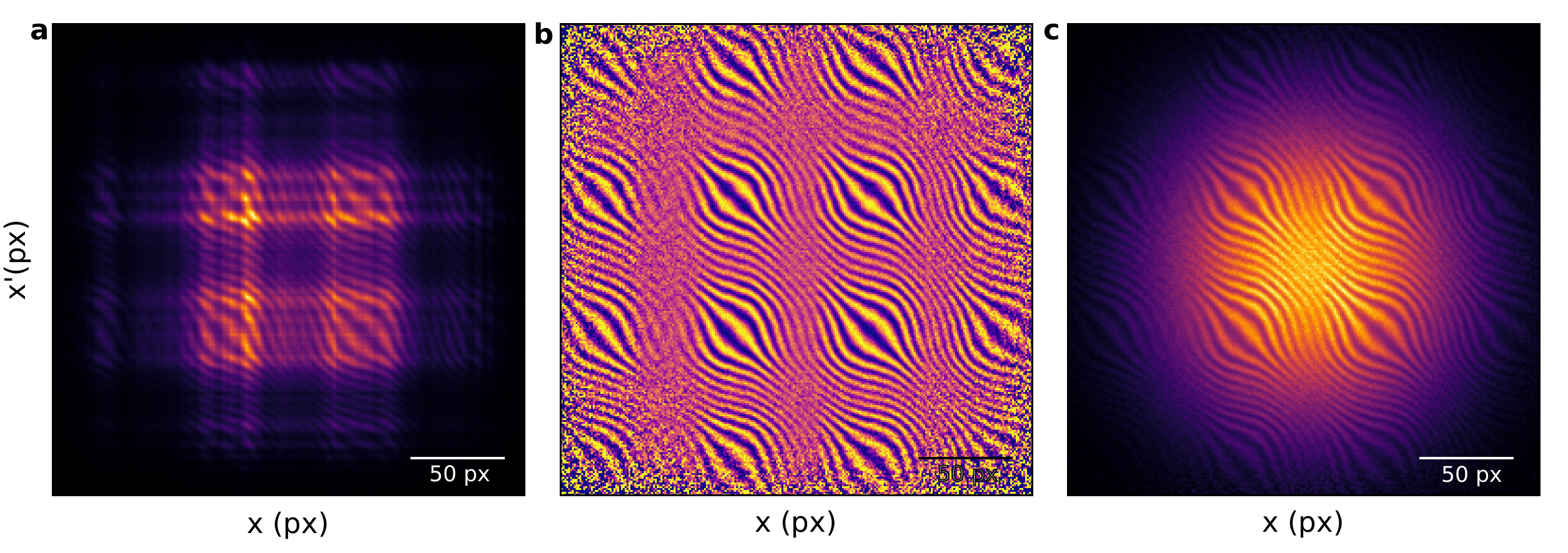}
   \caption[0.5\linewidth]{Additional steps for a highly distorted beam data processing: \textbf{a)} A raw coincidence map, created from all frames of the measurement. \textbf{b)} A normalized coincidence map in respect to the intensity of the beam. Even low photon count regions carry phase information.  \textbf{c)} normalized coincidence map from (b) multiplied by the Gaussian profile of the perfect beam. We add this step for the off-axis Fourier analysis to remove high-noise regions and smoothing the interferogram. Please note, that despite raw image being highly distorted from imperfections of the SLM, our phase retrieval works effectively and the normalized coincidence map has a high visibility.}
\label{fig:schem}
\end{figure}

\subsection*{S4.3 Quantitive error analysis}
We want to assess the accuracy of our phase imaging method in the most extreme scenario, in which only two photons are collected in each frame. Phase fluctuations are fast compared to the time between subsequent frames, but the fluctuating phase $\Theta $ is stable within a single frame---the same assumptions were made in S1 and S2.

Firstly, we need to construct the figure of merit assessing the phase retrieval accuracy. Let us assume, that our goal is to estimate the phase profile of the part of the object corresponding to $n$ pixels, whose numbers are $m,m+1,...,m+n-1$. The real phase in pixel $k$ is $\phi_k$, its estimation is $\tilde \phi_k$. Then, the MSE associated with the phase estimation in this region is
\begin{equation}
\label{eq:MSE}
    \textrm{MSE} = \left\langle \Delta^2 \phi_k \right\rangle = \frac{1}{n} \sum_{k=m}^{m+n-1} (\Delta \phi_k)^2= \frac{1}{n} \sum_{k=m}^{m+n-1} (\phi_k - \tilde \phi_k)^2,
\end{equation}
where the difference $\phi_k - \tilde \phi_k$ is always taken modulo $2 \pi$, such that $\phi_k - \tilde \phi_k \in [-\pi, \pi]$.

The theoretical minimum of the MSE can be obtained using C-R bound as described in S2---it is enough to insert Eq. \ref{pair} into Eq. \ref{eq:MSE}. This theoretical minimum is depicted in Figure 3 using solid, black line.

To check the optimality of our phase reconstruction method, we compare this minimum with the MSE obtained in the simulation. For a given beam profile $I_i$, phase profile $\phi_i$ and visibility $v$, we randomly sample $ n_\mathrm{phot}/2$ pairs of $x$ photons positions from the probability distribution described by Eq. \ref{diff}. Therefore, in our simulation, there are only two sources of noise---non-unit interference visibility and a shot noise, resulting from the finite number of photon pairs. The number of photon pairs is two times smaller than the total number of photons because one pair is created from each two-photon pair---this makes all pairs statistically independent. After creating an artificial hologram, we perform the whole phase reconstruction procedure described in \ref{Supp:analysis_fluct} to get phase estimators $\tilde \phi$. Then, we check a region of interest in the middle of the beam---we don't want to take into account regions where the beam intensity is very low. Finally, we compute the MSE associated with the phase estimation in this region using Eq. \ref{eq:MSE}. We repeat this procedure for different total number of photons $n_\mathrm{phot}$, which leads to different mean number of photons per pixel. We can observe in Figure 3, that the performance of our phase reconstruction method is optimal for large enough number of photons.

We also perform a similar, quantitative analysis using experimental data. To make the comparison between theory, experiment and simulation possible, we measure the beam intensity, phase profile and the visibility of 2nd order interference fringes in the correlation map ($v^2/2$). Then, the experimental parameters are used in the simulation and to calculate the C-R bound. We obtain phase estimators $\tilde \phi_i$ using experimentally measured correlation map. To obtain such a map, we randomly choose just one photon pair from each frame---the goal is to simulate the experimental conditions, in which only two photons per frame are available. The ground truth phase profile $\phi_i$ is quadratic because we use a standard, thin lens as a phase mask. The parameters of this ground truth phase profile are measured using the standard technique with stable phases in both arms. The experimental phase estimators $\tilde \phi_i$ are calculated without any assumptions about the shape of the phase profile---we don't assume, that the measured phase profile is quadratic. The MSE for the region of interest is again computed using Eq. \ref{eq:MSE}. We observe, that the MSE from experiment and simulation are very similar- they only start to differ slightly for the number of photons per pixel $\sim 4 \times 10^4$. This may suggest, that noise sources other than shot noise start to play a significant role at this point.

\bibliography{scibib}

\bibliographystyle{Science.bst}